\documentstyle[preprint,eqsecnum,tighten,aps]{revtex}
\makeatletter
\def\references{%
\ifpreprintsty
\vskip 0.8cm plus1ex minus .2ex
\hbox to\hsize{\hss\bf REFERENCES \hss}%
\else
\vskip24pt
\hrule width\hsize\relax
\vskip 1.6cm
\fi
\list{\@biblabel{\arabic{enumiv}}}%
{\labelwidth\WidestRefLabelThusFar  \labelsep4pt %
\leftmargin\labelwidth %
\advance\leftmargin\labelsep %
\ifdim\baselinestretch pt>1 pt %
\parsep  4pt\relax %
\else %
\parsep  0pt\relax %
\fi
\itemsep\parsep %
\usecounter{enumiv}%
\let\p@enumiv\@empty
\def\theenumiv{\arabic{enumiv}}%
}%
\let\newblock\relax %
\sloppy\clubpenalty4000\widowpenalty4000
\sfcode`\.=1000\relax
\ifpreprintsty\else\small\fi
}
\def\figure{%
\let\@capwidth\columnwidth
\ifpreprintsty\iffirstfig
{\vskip 0.8cm plus1ex minus .2ex
\centerline{\bf FIGURES}}\global\firstfigfalse
\fi\fi
\vskip1pc
\def\@captype{figure}%
\interlinepenalty10000 %
\@ifnextchar[{\@chuckoptarg}{}%
}%
\def\table{%
\let\@capwidth\columnwidth \def\@tablenotes{}%
\iffirsttab
\global\firsttabfalse
\ifpreprintsty{\newpage\centerline{\bf TABLES}}\fi
\fi
\vskip1pc
\global\tableontrue
\bgroup\parindent=0pt
\outertabtrue
\setcounter{tablenote}{0}%
\def\@captype{table}%
\@ifnextchar[{\@chuckoptarg}{}%
}%
\def\endtable{%
\global\tableonfalse\global\outertabfalse
{\let\protect\relax\small\vskip2pt\@tablenotes\par}\xdef\@tablenotes{}%
\egroup
}%
\makeatother
\begin{document}
\draft
\title{Measurement of the $ZZ\gamma$ and $Z\gamma\gamma$ Couplings
in $p\bar p$ Collisions at $\sqrt{s} = 1.8$~TeV}
%
% LIST_OF_AUTHORS.TEX                 03/09/95
%
\author{
%% names begin here
S.~Abachi,$^{12}$
B.~Abbott,$^{33}$
M.~Abolins,$^{23}$
B.S.~Acharya,$^{40}$
I.~Adam,$^{10}$
D.L.~Adams,$^{34}$
M.~Adams,$^{15}$
S.~Ahn,$^{12}$
H.~Aihara,$^{20}$
J.~Alitti,$^{36}$
% ...
\\
\vskip 0.50cm
\centerline{(D\O\ Collaboration)}
\vskip 0.50cm
}
\address{
\centerline{$^{1}$Universidad de los Andes, Bogota, Colombia}
\centerline{$^{2}$University of Arizona, Tucson, Arizona 85721}
\centerline{$^{3}$Brookhaven National Laboratory, Upton, New York 11973}
\centerline{$^{4}$Brown University, Providence, Rhode Island 02912}
\centerline{$^{5}$University of California, Davis, California 95616}
\centerline{$^{6}$University of California, Irvine, California 92717}
\centerline{$^{7}$University of California, Riverside, California 92521}
\centerline{$^{8}$LAFEX, Centro Brasileiro de Pesquisas F{\'\i}sicas,
                  Rio de Janeiro, Brazil}
\centerline{$^{9}$CINVESTAV, Mexico City, Mexico}
\centerline{$^{10}$Columbia University, New York, New York 10027}
\centerline{$^{11}$Delhi University, Delhi, India 110007}
\centerline{$^{12}$Fermi National Accelerator Laboratory, Batavia,
                   Illinois 60510}
\centerline{$^{13}$Florida State University, Tallahassee, Florida 32306}
\centerline{$^{14}$University of Hawaii, Honolulu, Hawaii 96822}
\centerline{$^{15}$University of Illinois, Chicago, Illinois 60680}
\centerline{$^{16}$Indiana University, Bloomington, Indiana 47405}
\centerline{$^{17}$Iowa State University, Ames, Iowa 50011}
\centerline{$^{18}$Korea University, Seoul, Korea}
\centerline{$^{19}$Kyungsung University, Pusan, Korea}
\centerline{$^{20}$Lawrence Berkeley Laboratory, Berkeley, California 94720}
\centerline{$^{21}$University of Maryland, College Park, Maryland 20742}
\centerline{$^{22}$University of Michigan, Ann Arbor, Michigan 48109}
\centerline{$^{23}$Michigan State University, East Lansing, Michigan 48824}
\centerline{$^{24}$Moscow State University, Moscow, Russia}
\centerline{$^{25}$University of Nebraska, Lincoln, Nebraska 68588}
\centerline{$^{26}$New York University, New York, New York 10003}
\centerline{$^{27}$Northeastern University, Boston, Massachusetts 02115}
\centerline{$^{28}$Northern Illinois University, DeKalb, Illinois 60115}
\centerline{$^{29}$Northwestern University, Evanston, Illinois 60208}
\centerline{$^{30}$University of Notre Dame, Notre Dame, Indiana 46556}
\centerline{$^{31}$University of Panjab, Chandigarh 16-00-14, India}
\centerline{$^{32}$Institute for High Energy Physics, 142-284 Protvino, Russia}
\centerline{$^{33}$Purdue University, West Lafayette, Indiana 47907}
\centerline{$^{34}$Rice University, Houston, Texas 77251}
\centerline{$^{35}$University of Rochester, Rochester, New York 14627}
\centerline{$^{36}$CEA, DAPNIA/Service de Physique des Particules, CE-SACLAY,
                   France}
\centerline{$^{37}$Seoul National University, Seoul, Korea}
\centerline{$^{38}$State University of New York, Stony Brook, New York 11794}
\centerline{$^{39}$SSC Laboratory, Dallas, Texas 75237}
\centerline{$^{40}$Tata Institute of Fundamental Research,
                   Colaba, Bombay 400005, India}
\centerline{$^{41}$University of Texas, Arlington, Texas 76019}
\centerline{$^{42}$Texas A\&M University, College Station, Texas 77843}
}
\maketitle
\vspace{-0.2in}
\begin{abstract}
We have directly  measured the $ZZ\gamma$ and  $Z\gamma\gamma$ couplings by
studying   $p\bar p \to   \ell\ell\gamma  + X$,   $(\ell=e,\mu)$  events at
$\sqrt{s}=1.8$~TeV    with  the D\O\   detector at  the  Fermilab  Tevatron
Collider. A  fit to the  transverse  energy  spectrum of the  photon in the
signal  events,  based  on the  data set   corresponding  to an  integrated
luminosity  of  $13.9\  {\rm  pb}^{-1}$  ($13.3\  {\rm   pb}^{-1}$) for the
electron  (muon)  channel,  yields the  following  $95\%$  confidence level
limits  on the   anomalous ${\it    CP}$-conserving  $ZZ\gamma$  couplings:
$-1.9<h^Z_{30}<1.8$ ($h^Z_{40} = 0$), and $-0.5<h^Z_{40}<0.5$ ($h^Z_{30}$ =
0),  for a    form-factor  scale   $\Lambda =    500$~GeV.  Limits  for the
$Z\gamma\gamma$   couplings and  ${\it   CP}$-violating  couplings are also
discussed.
\end{abstract}
\bigskip
\pacs{\it Submitted to Physical Review Letters}

\section{Introduction}

Direct  measurement of the  $ZZ\gamma$ and  $Z\gamma\gamma$ trilinear gauge
boson couplings is  possible by studying  $Z\gamma$ production in $p\bar p$
collisions at the Tevatron ($\sqrt{\it{s}}=1.8$ TeV). In what follows these
couplings will  be addressed  to as  $ZV\gamma$, where $V =  Z,\gamma$. The
most general Lorentz and gauge  invariant $ZV\gamma$ vertex is described by
four coupling  parameters,  $h^V_i,~(i=1...4)$~\cite{ref1}. Combinations of
the ${\it  CP}$-conserving (${\it  CP}$-violating)  parameters  $h^V_3$ and
$h^V_4$ ($h^V_1$ and $h^V_2$) correspond to the electric  (magnetic) dipole
and magnetic  (electric)  quadrupole  transition moments of  the $ZV\gamma$
vertex. In the Standard Model (SM),  all the $ZV\gamma$ couplings vanish at
the tree  level.  Non-zero  (i.e. {\it  anomalous})  values of  the $h^V_i$
couplings result in an  increase of the $Z\gamma$  production cross section
and    change   the    kinematic      distribution  of  the    final  state
particles~\cite{ref2}.  Partial wave unitarity of  the general $f\bar f \to
Z\gamma$  process  restricts  the  $ZV\gamma$  couplings  uniquely to their
vanishing  SM  values at   asymptotically high   energies~\cite{unitarity}.
Therefore, the   coupling  parameters have  to be modified  by form-factors
$h^V_i =  h^V_{i0} / (1  +   \hat{s}/\Lambda^2)^n$, where  $\hat{s}$ is the
square of  the invariant  mass of  the $Z\gamma$  system,  $\Lambda$ is the
form-factor  scale, and  $h^V_{i0}$ are  coupling values at  the low energy
limit   ($\hat{s}\approx   0$)~\cite{ref2}.  Following  Ref.~\cite{ref2} we
assume $n = 3$ for  $h^V_{1,3}$ and $n = 4$ for  $h^V_{2,4}$. Such a choice
yields the same  asymptotic energy  behavior for all  the couplings. Unlike
$W\gamma$  production where the  form-factor effects do  not play a crucial
role,  the   $\Lambda$-dependent  effects  cannot be  ignored in  $Z\gamma$
production  at  Tevatron   energies.  This is due  to the  higher  power of
$\hat{s}$ in the  vertex function, a  direct  consequence of the additional
Bose-Einstein symmetry of the $ZV\gamma$ vertices~\cite{ref2}.

We present a  measurement of  the  $ZV\gamma$ couplings  using $p\bar p \to
\ell\ell\gamma +  X~ (\ell=e,\mu)$  events observed  with the D\O\ detector
during the  1992--1993 run,  corresponding  to an  integrated luminosity of
$13.9 \pm 1.7 {\rm pb}^{-1}$ ($13.3\pm 1.6 {\rm pb}^{-1}$) for the electron
(muon) data. Similar measurements were recently performed by CDF~\cite{CDF}
and L3~\cite{L3}.

\section{Detector and Event Selection}

The D\O\ detector,  described in detail  elsewhere~\cite{ref3}, consists of
three  main  systems. The   calorimeter  consists of   uranium-liquid argon
sampling  detectors  in the  central and  two end  cryostats,  and provides
near-hermetic coverage in pseudorapidity ($\eta$) for $|\eta|\leq 4.4.$ The
energy    resolution  of  the    calorimeter  has  been   measured  in beam
tests~\cite{ref4} to be  $15\%/\sqrt{E}$ for  electrons and $50\%/\sqrt{E}$
for isolated  pions, where  $E$ is in GeV.  The calorimeter  is read out in
towers that subtend $0.1\times 0.1$  in $\eta \times \phi$ (where $\phi$ is
the    azimuthal   angle)  and  are    segmented     longitudinally  into 4
electromagnetic (EM) and $4 - 5$ hadronic layers. In the third EM layer, at
the EM shower  maximum, the towers  are more finely  subdivided, subtending
$0.05\times 0.05$ in  $\eta\times\phi$. Central  and forward drift chambers
are used to identify  charged tracks for  $|\eta|\leq 3.2.$ The muon system
consists of magnetized iron toroids  with one inner and two outer layers of
drift tubes,  providing  coverage for  $|\eta|\leq 3.3.$  The muon momentum
resolution for central  muons ($|\eta| < 1.0$) is  determined to be $\delta
(1/p)/(1/p) =  0.18(p-2)/p  \oplus 0.008 p$  ($p$ in  GeV$/c$), using $Z\to
\mu\mu$ events.

$Z\gamma$  candidates are selected  by searching for  events containing two
isolated  electrons (muons) with  high transverse  energy $E_T$ (transverse
momentum $p_T$), and an isolated  photon. The $ee\gamma$ sample is selected
from a  trigger  requiring two  isolated EM  clusters, each  with $E_T \geq
20$~GeV. An electron  cluster is required to be  within the fiducial region
of the  calorimeter ($|\eta|\leq  1.1$ in the central  calorimeter (CC), or
$1.5\leq  |\eta|\leq 2.5$ in the end  calorimeters  (EC)). Offline electron
identification requirements are:  ({\it i\/}) the ratio of the EM energy to
the total  shower  energy must  be $>  0.9$; ({\it  ii\/}) the  lateral and
longitudinal     shower   shape  must  be   consistent   with an   electron
shower~\cite{meena};  ({\it iii\/})  the isolation  variable of the cluster
(${\it  I}$) must be  $< 0.1$,  where  ${\it I}$ is  defined as  ${\it I} =
[E_{\rm tot}(0.4) -  E_{\rm EM}(0.2)]/E_{\rm  EM}(0.2)$, $E_{\rm tot}(0.4)$
is  the   total   shower    energy   inside  a  cone    defined  by  ${\cal
R}=\sqrt{(\Delta\eta)^2+(\Delta\phi)^2}=0.4$,  and $E_{\rm EM}(0.2)$ is the
EM energy inside a cone of ${\cal R}=0.2$; ({\it iv\/}) at least one of the
two electron  clusters must  have a matching  track in the  drift chambers;
and ({\it v\/}) $E_T > 25$~GeV for both electrons.

The   $\mu\mu\gamma$  sample is  selected  from a  trigger  requiring an EM
cluster with $E_T > 7$~GeV and a  muon track with $p_T > 5$~GeV$/c$. A muon
track is required to have $|\eta| \leq 1.0$ and must have: ({\it i\/}) hits
in the inner drift-tube layer; ({\it ii\/}) a good overall track fit; ({\it
iii\/}) bend view impact parameter $< 22$~cm; ({\it iv\/}) a matching track
in the central drift chambers; and ({\it v\/}) minimum energy deposition of
1~GeV in the  calorimeter  along the muon  path. The muon  must be isolated
from a nearby  jet $({\cal  R}_{\mu-{\rm  jet}}>0.5)$. At  least one of the
muon tracks is  required to  traverse a  minimum length of  magnetized iron
($\int  Bdl  >   1.9$~Tm); it  is  also  required  that    $p_T^{\mu_{1}} >
15$~GeV$/c$ and $p_T^{\mu_{2}} > 8$~GeV$/c$.

The requirements for photon  identification are common to both electron and
muon samples. We require  a photon transverse  energy $E^\gamma_T > 10$~GeV
and the same quality cuts as those  on the electron, except that there must
be no  track  pointing toward  the  calorimeter  cluster.  Additionally, we
require that the  separation between  a photon and both  leptons be $\Delta
{\cal R}_{\ell\gamma} >  0.7$. This cut  suppresses the contribution of the
radiative $Z \to  \ell\ell\gamma$  decays~\cite{ref2}.  The above selection
criteria   yield  four   $ee\gamma$ and  two    $\mu\mu\gamma$  candidates.
Figure~\ref{fig:Et}  shows the $E_T^\gamma$  distribution for these events.
Three  $ee\gamma$  and both   $\mu\mu\gamma$  candidates have a  three body
invariant  mass close to  that of  the $Z$ and  low  separation between the
photon and one of the leptons,  consistent with the interpretation of these
events  as  radiative $Z  \to  \ell\ell  \to   \ell\ell\gamma$  decays. The
remaining  candidate in electron  channel has a  dielectron mass compatible
with that of the $Z$ and a photon well separated from the leptons, an event
topology  typical  for direct  $Z\gamma$  production  in which  a photon is
radiated from one of the interacting partons~\cite{ref2}.

\section{Signal and Backgrounds}

The estimated  background,  summarized in  Table~I,  includes contributions
from ({\it i\/}) $Z+ {\rm jet(s)}$ production where one of the jets fakes a
photon  or an  electron  (the latter  case  corresponds  to the  $ee\gamma$
signature if  additionally one of the electrons  from the $Z \to e e$ decay
is  not  detected  in a  tracking   chamber);  ({\it  ii\/})  QCD  multijet
production  with jets being  misidentified  as electrons or  photons; ({\it
iii\/})  $\tau\tau\gamma$  production  followed by decay of  each $\tau$ to
$\ell\overline\nu_\ell\nu_\tau$.

We estimate  the QCD  background from data  using the  probability, $P({\rm
jet} \to e/\gamma)$, for  a jet to be  misidentified as an electron/photon.
This  probability is  determined by  measuring the fraction  of non-leading
jets in  samples  of QCD  multijet  events  that pass  our  photon/electron
identification cuts, and  takes into account a  $0.25 \pm 0.25$ fraction of
direct  photon events in  the  multijet   sample~\cite{direct}. We find the
misidentification   probabilities $P({\rm  jet} \to  e/\gamma)$ to be $\sim
10^{-3}$  in the  typical $E_T$  ranges  for the  electrons and  photons of
between 10 and  50~GeV. We find the  background from  $Z+ {\rm jet(s)}$ and
QCD multijet events in  the electron channel by  applying misidentification
probabilities   to the  jet  $E_T$  spectrum of  the  inclusive  $ee + {\rm
jet(s)}$ and  $e\gamma + {\rm  jet(s)}$  data. The  background is $0.43 \pm
0.06$  events. For  the muon  channel the  QCD  background is  estimated by
applying the misidentification  probability to the inclusive $\mu\mu + {\rm
jet(s)}$  spectrum. The estimation  of the QCD  background from data in the
muon case also accounts for cosmic  ray background. The combined background
from QCD multijet and cosmic ray  events in the muon channel is found to be
$0.02 \pm 0.01$ events.

The $\tau\tau\gamma$  background is estimated  using the ISAJET Monte Carlo
event   generator~\cite{ref7}  followed  by a full  simulation  of the D\O\
detector,    resulting in   $0.004\pm  0.002$  events  for   $ee\gamma$ and
$0.03\pm0.01$ events for $\mu\mu\gamma$ channels.

Subtracting the  estimated backgrounds from the  observed number of events,
the signal is $3.57^{+3.15}_{-1.91}\pm 0.06$ for the $ee\gamma$ channel and
$1.95^{+2.62}_{-1.29}\pm  0.01$ for  the $\mu\mu\gamma$  channel, where the
first and dominant uncertainty is due to Poisson statistics, and the second
is due to the systematic error of the background estimate.

The acceptance of the  D\O\ detector for the  $ee\gamma$ and $\mu\mu\gamma$
final states was  studied using the  leading order  event generator of Baur
and Berger \cite{ref2}. It generates  4-vectors for the $Z\gamma$ processes
as a function of the coupling parameters. The 4-vectors were then processed
using a fast detector simulation  program which takes into account  effects
of the  electromagnetic  and missing  transverse  energy  resolutions, muon
momentum   resolution,   variations in  position  of the  vertex  along the
beam-axis, and  trigger and  offline  efficiencies. These  efficiencies are
estimated using $Z\to  ee$ data for the electron  channel. The muon trigger
efficiency  is  estimated  from the  $e\mu$  data  selected  using non-muon
triggers. The offline  efficiency for the muon  channel is calculated based
on $e\mu$ and  $Z\to\mu\mu$ samples. The trigger  efficiency for $ee\gamma$
is  $0.98  \pm   0.01$  while  the    efficiency  of   offline   dielectron
identification is $0.64  \pm0.02$ in the CC and  $0.56 \pm 0.03$ in the EC.
For the muon channel the trigger  efficiency is $0.94^{+0.06}_{-0.09}$, and
the offline dimuon identification efficiency is $0.54 \pm 0.04$. The photon
efficiency depends slightly on  $E_T^\gamma$ due to the calorimeter cluster
shape algorithm and the isolation  cut, and accounts for loss of the photon
due to a random  track overlap  (which results in  misidentification of the
photon as  an electron)  and the  photon  conversion into an  $e^+e^-$ pair
before the  outermost  tracking chamber. The  average  photon efficiency is
$0.53 \pm 0.05$. The geometrical acceptance for the electron (muon) channel
is  53\%  (20\%) for  the SM  case  and  increases   slightly for  non-zero
anomalous couplings. The overall efficiency for the electron (muon) channel
for  SM   couplings is  $0.17  \pm  0.02$  ($0.06  \pm   0.01$).  The ${\rm
MRSD-}'$~\cite{MRSD}      set  of   structure   functions  is  used  in the
calculations.  The  uncertainties due to the  choice of  structure function
(6\%, as  determined by  variation of the  results for  different sets) are
included in the systematic error of the Monte Carlo calculation. The effect
of higher order QCD corrections are  accounted for by multiplying the rates
by a constant factor $k = 1.34$~\cite{ref2}.

The observed number of events is compared with the SM expectation using the
estimated  efficiency and  acceptance. We expect the  signal in the $e$ and
$\mu$ channels for SM  couplings to be:  $S_{ee\gamma}^{\it SM}$ = $2.7 \pm
0.3~\mbox{(sys)}  \pm  0.3~\mbox{(lum)}$ and  $S_{\mu\mu\gamma}^{\it SM}$ =
$2.2 \pm  0.4~\mbox{(sys)}  \pm  0.3~\mbox{(lum)}$ events,  where the first
error is  due to the   uncertainty in the  Monte  Carlo  modelling, and the
second reflects the  uncertainty in the  integrated luminosity calculation.
The numbers are  summarized in  Table~I. Our observed  signal agrees within
errors with the SM prediction for both channels.

\section{Limits on Anomalous Couplings}

\begin{figure}[bth]
\vspace*{4.2in}
\includegraphics{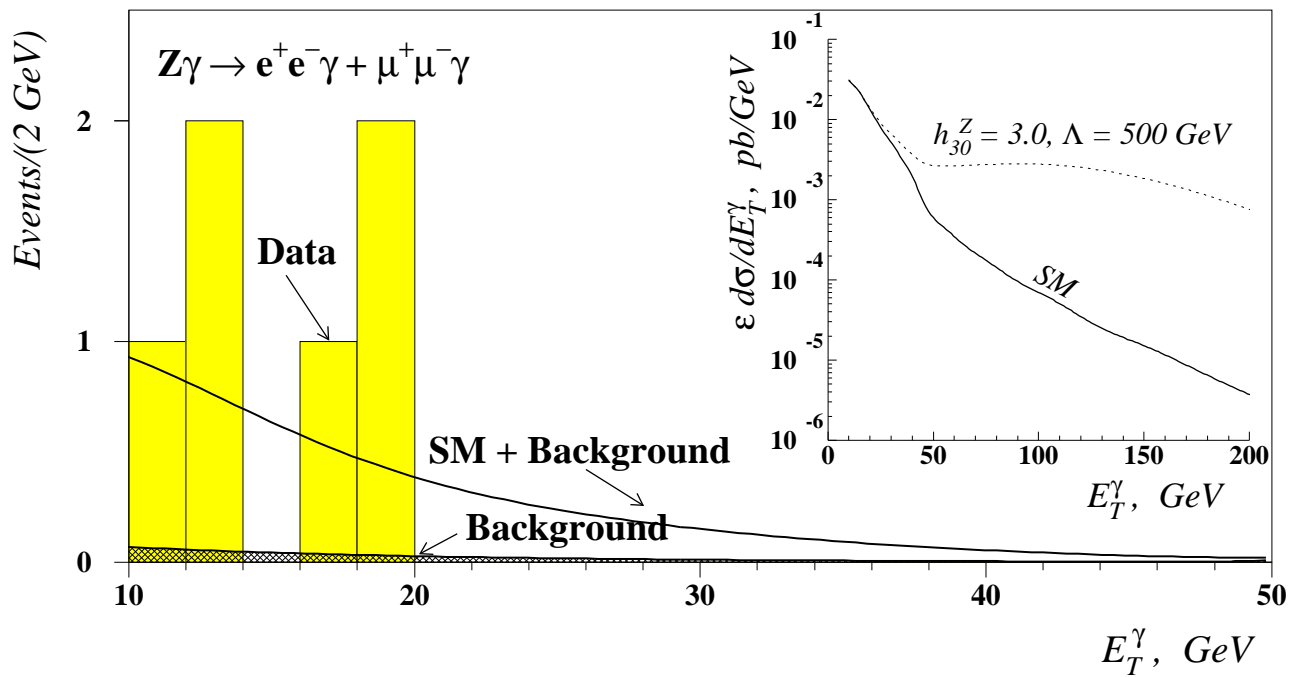}
\caption{Transverse      energy  spectrum  of  photons  in   $ee\gamma$ and
$\mu\mu\gamma$  events. The  shadowed bars  correspond to  data points, the
hatched curve represents the total background, and the solid line shows the
sum  of  the SM    predictions  and  the    background.  The  insert  shows
$d\sigma/dE_T^\gamma$  folded  with the  efficiencies for  SM and anomalous
($h_{30}^Z = 3.0$)  couplings.}
\label{fig:Et}
\end{figure}

To set limits  on the  anomalous coupling  parameters, we  fit the observed
$E_T$ spectrum  of the photon   ($\it{E_T^{\gamma}}$) with  the Monte Carlo
predictions plus the estimated background, combining the information in the
spectrum shape and the event rate.  The fit is performed for the $ee\gamma$
and $\mu\mu\gamma$  samples, using a binned  likelihood method~\cite{Greg},
including  constraints to account  for our  understanding of luminosity and
efficiency    uncertainties.  Because  the  contribution  of the  anomalous
couplings is concentrated in the high $E_T^\gamma$ region, the differential
distribution    $d\sigma/dE_T^\gamma$ is  more  sensitive to  the anomalous
couplings than a total  cross section (see insert  in Fig.~\ref{fig:Et} and
Ref.~\cite{ref2}).  To optimize the  sensitivity of the  experiment for the
low statistics, we assume Poisson  statistics for each $E_T^\gamma$ bin and
use the maximum likelihood method to  fit the experimental data. To exploit
the fact that anomalous coupling  contributions lead to an excess of events
at high transverse energy of the  photon, a high-$E_T^\gamma$ bin, in which
we observe no events is  explicitly used in the  histogram~\cite{Greg}. The
results were  cross-checked using an  unbinned  likelihood fit which yields
similar results.

\begin{figure}
\vspace*{7.3in}
\includegraphics{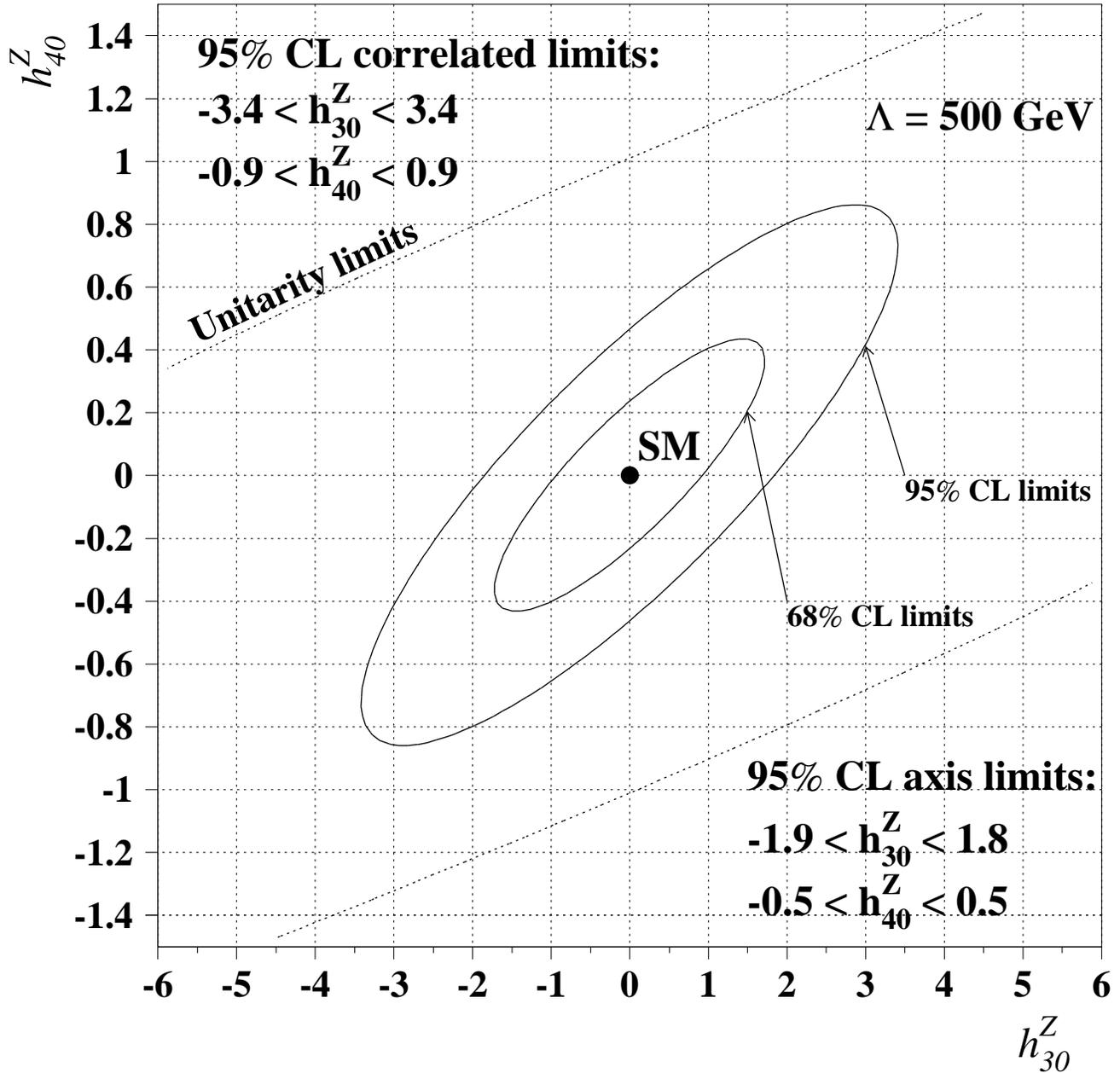}
\caption{Limits on the  correlated ${\it  CP}$-con\-ser\-ving ano\-ma\-lous
$ZZ\gamma$   coupling   parameters  $h^Z_{30}$  and  $h^Z_{40}$.  The solid
ellipses  represent 68\% and 95\% CL  exclusion  contours. The dashed curve
shows limits from partial wave unitarity for $\Lambda = 500$~GeV.}
\label{fig:hz34}
\end{figure}

Figure~\ref{fig:Et}  shows the  observed  $\it{E_T^{\gamma}}$ spectrum with
the SM prediction  plus the  estimated background for  the $e+\mu$ combined
sample.  The  95\%   confidence  level (CL)  limit  contour  for  the ${\it
CP}$-conserving anomalous coupling  parameters $h^Z_{30}$ and $h^Z_{40}$ is
shown in  Fig.~\ref{fig:hz34}. A  form-factor scale  $\Lambda = 500$~GeV is
used for  the  calculations of  the  experimental  limits and  partial wave
unitarity constraints. We obtain the following 95\% CL limits for the ${\it
CP}$-conserving $ZZ\gamma$ and $Z\gamma\gamma$ couplings (in the assumption
that all couplings except one are at the SM values, i.e. zeros):

{\centerline{$-1.9<h^Z_{30}<1.8$; $-0.5 < h^Z_{40} < 0.5$}}

{\centerline{$-1.9<h^\gamma_{30}<1.9$; $-0.5<h^\gamma_{40}<0.5$}}

\noindent
The correlated limits for pairs of couplings $(h^V_{30},h^V_{40})$ are less
stringent due to the strong  interference between these couplings:

{\centerline{$-3.3<h^Z_{30}<3.3$; $-0.9 < h^Z_{40} < 0.9$}}

{\centerline{$-3.5<h^\gamma_{30}<3.5$; $-0.9 < h^\gamma_{40} < 0.9$}}

\noindent
Limits on the ${\it CP}$-violating $ZV\gamma$ couplings are numerically the
same as those for  the ${\it  CP}$-conserving  couplings. The limits on the
$h^Z_{20}$,  $h^Z_{40}$, and   $h^\gamma_{i0}$ couplings  are currently the
most stringent.

\begin{figure}
\vspace*{7.3in}
\includegraphics{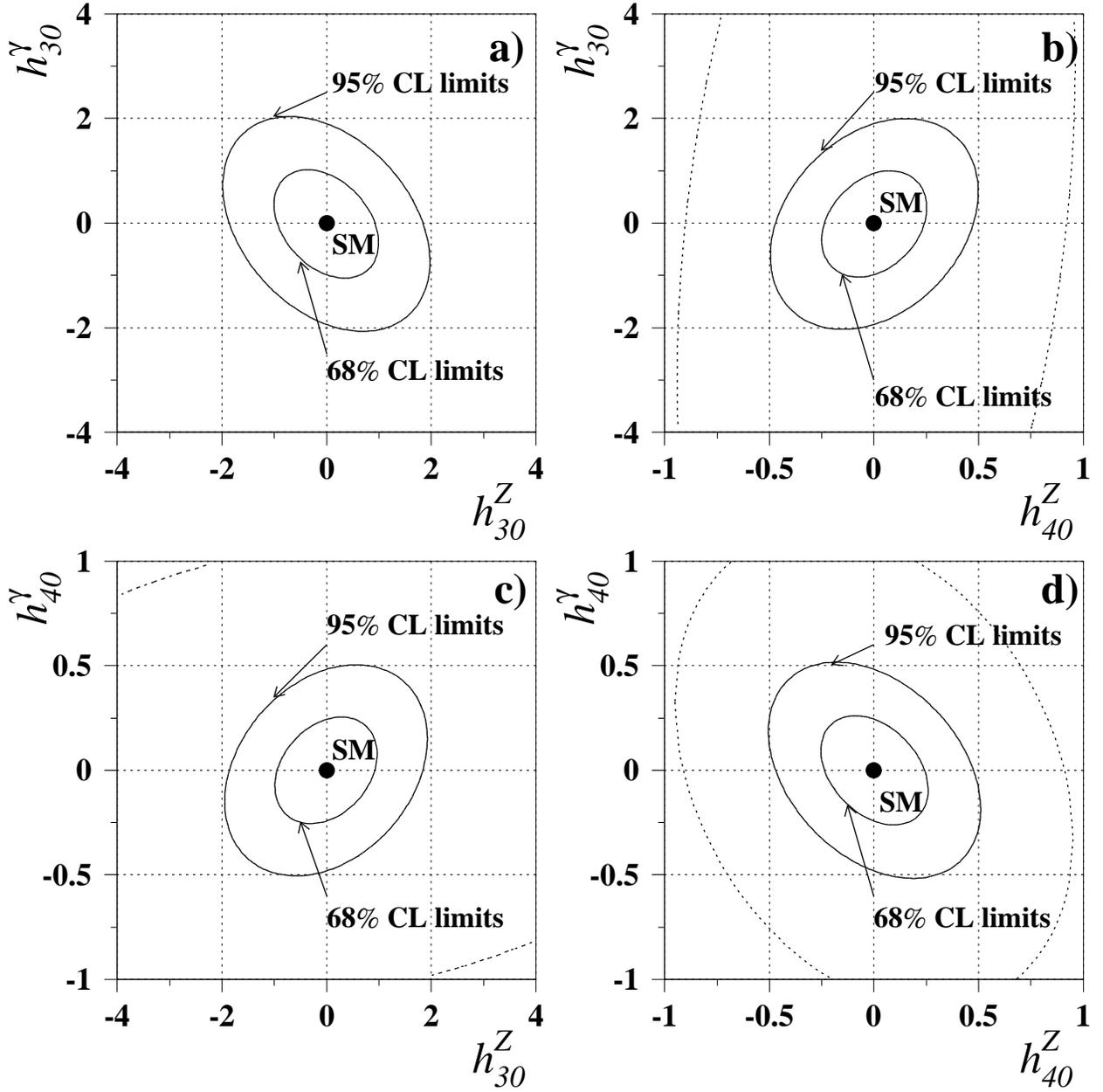}
\caption{Limits  on the  weakly correlated  ${\it  CP}$-conserving pairs of
anomalous    $ZV\gamma$   couplings: a)      $(h^Z_{30},h^\gamma_{30})$, b)
$(h^Z_{40},h^\gamma_{30})$,      c)     $(h^Z_{30},h^\gamma_{40})$,  and d)
$(h^Z_{40},h^\gamma_{40})$.  The solid ellipses  represent 68\% and 95\% CL
exclusion contours.  Dashed curves show limits  from partial wave unitarity
for $\Lambda = 500$~GeV.}
\label{fig:cross}
\end{figure}

Global limits on the  anomalous couplings (i.e.,  limits independent of the
values  of  other   couplings)  are  close to  the   correlated  limits for
$(h^V_{30},h^V_{40})$ and $(h^V_{10},h^V_{20})$ pairs, since other possible
combinations of  couplings interfere  with each other  only at the level of
10\%. This is  illustrated in  Fig.~\ref{fig:cross}, which shows the limits
for  pairs of  couplings of  the  same ${\it   CP}$-parity  (couplings with
different ${\it CP}$-parity do not interfere with each other).

We also study the  form-factor scale dependence  of the results. The chosen
value of the scale $\Lambda = 500$~GeV is close to the sensitivity limit of
this experiment for the  $h^V_{20,40}$ couplings:  for larger values of the
scale partial  wave unitarity is  violated for certain  values of anomalous
couplings allowed at 95\% CL by this measurement.
\newpage

\begin{acknowledgements}

We would like to  thank U.~Baur for  providing us with  the $Z\gamma$ Monte
Carlo program and for many helpful  discussions. We also thank the Fermilab
Accelerator,  Computing and Research  Divisions, and  the support staffs at
the collaborating  institutions for  their  contributions to the success of
this  work. We  also  acknowledge  the  support of the  U.S.  Department of
Energy,  the  U.S.  National   Science  Foundation,  the   Commissariat \`a
L'Energie   Atomique in  France, the   Ministry for  Atomic  Energy and the
Ministry of  Science and Technology  Policy in Russia,  CNPq in Brazil, the
Departments   of  Atomic   Energy  and  Science  and   Education in  India,
Colciencias  in  Colombia, CONACyT  in Mexico,  the Ministry  of Education,
Research Foundation and KOSEF in Korea and the A.P. Sloan Foundation.
\end{acknowledgements}

\begin{table}[p]
\vspace{0.3in}
\label{table1}
\begin{tabular}{lcc}
 & $ee\gamma$ & $\mu\mu\gamma$ \\
\tableline
Candidates & 4 & 2 \\
\tableline
Background: & & \\
QCD & $0.43 \pm 0.06$ & $0.02 \pm 0.01$ \\
$\tau\tau\gamma$ & $0.004\pm 0.002$ & $0.03\pm0.01$ \\
\tableline
%                   &                       &                        \\
Total background        & $0.43 \pm 0.06 $ & $0.05 \pm 0.01$ \\
\tableline
%                   &                       &                        \\
Signal\rule[-0.1in]{0pt}{0.3in}         & $3.57^{+3.15}_{-1.91}
\pm 0.06$ & $1.95^{+2.62}_{-1.29}\pm 0.01$\\
\tableline
%                   &                       &                        \\
SM predictions     & $2.7 \pm 0.3 \pm 0.3$  & $2.3 \pm 0.4 \pm 0.3$\\
\end{tabular}
\vspace{0.1in}
\caption{Summary of signal and backgrounds.}
\vspace{0.2in}
\end{table}

\end{document}